\documentstyle{mn}

\newif\ifAMStwofonts

\def\Hip{{\it Hipparcos\/}}
\def\Msun{\mbox{$\,{\rm M}_\odot$}}
\def\parc{\mbox{$\,{\rm pc}$}}
\def\Mden{\ifmmode{~{\rm M}_\odot~{\rm pc}^{-3}}
 \else${\rm M}_\odot$~pc$^{-3}$\fi}

\ifoldfss
  \ifCUPmtlplainloaded \else
    \NewTextAlphabet{textbfit} {cmbxti10} {}
    \NewTextAlphabet{textbfss} {cmssbx10} {}
    \NewMathAlphabet{mathbfit} {cmbxti10} {} 
    \NewMathAlphabet{mathbfss} {cmssbx10} {} 
  \fi
  \ifAMStwofonts
    \ifCUPmtlplainloaded \else
      \NewSymbolFont{upmath} {eurm10}
      \NewSymbolFont{AMSa} {msam10}
      \NewMathSymbol{\upi}     {0}{upmath}{19}
      \NewMathSymbol{\umu}     {0}{upmath}{16}
      \NewMathSymbol{\upartial}{0}{upmath}{40}
      \NewMathSymbol{\leqslant}{3}{AMSa}{36}
      \NewMathSymbol{\geqslant}{3}{AMSa}{3E}

      \let\leq=\leqslant \let\le=\leqslant
       
    \fi
  \fi
\fi 

\ifnfssone
  \newmathalphabet{\mathit}
  \addtoversion{normal}{\mathit}{cmr}{m}{it}
  \addtoversion{bold}{\mathit}{cmr}{bx}{it}
  \def\textbfit{\protect\txtbfit}
  
  \long\def\txtbfit#1{{\fontfamily{cmr}\fontseries{bx}\fontshape{it}%
    \selectfont #1}}

  \newmathalphabet{\mathbfit} 
  \addtoversion{normal}{\mathbfit}{cmr}{bx}{it}
  \addtoversion{bold}{\mathbfit}{cmr}{bx}{it}
  \newmathalphabet{\mathbfss} 
  \addtoversion{normal}{\mathbfss}{cmss}{bx}{n}
  \addtoversion{bold}{\mathbfss}{cmss}{bx}{n}
  \ifAMStwofonts
    \ifCUPmtlplainloaded \else
      \UseAMStwoboldmath
      \makeatletter
      \new@mathgroup\upmath@group
      \define@mathgroup\mv@normal\upmath@group{eur}{m}{n}
      \define@mathgroup\mv@bold\upmath@group{eur}{b}{n}
      \edef\UPM{\hexnumber\upmath@group}
      \new@mathgroup\amsa@group
      \define@mathgroup\mv@normal\amsa@group{msa}{m}{n}
      \define@mathgroup\mv@bold\amsa@group{msa}{m}{n}
      \edef\AMSa{\hexnumber\amsa@group}
      \makeatother
      \mathchardef\upi="0\UPM19
      \mathchardef\umu="0\UPM16
      \mathchardef\upartial="0\UPM40
      \mathchardef\leqslant="3\AMSa36
      \mathchardef\geqslant="3\AMSa3E

      \let\leq=\leqslant \let\le=\leqslant

    \fi
  \fi
\fi 

\ifnfsstwo
  \def\textbfit{\protect\txtbfit}
  
  \long\def\txtbfit#1{{\fontfamily{cmr}\fontseries{bx}\fontshape{it}%
    \selectfont #1}}

  \DeclareMathAlphabet{\mathbfit}{OT1}{cmr}{bx}{it}
  \SetMathAlphabet\mathbfit{bold}{OT1}{cmr}{bx}{it}
  \DeclareMathAlphabet{\mathbfss}{OT1}{cmss}{bx}{n}
  \SetMathAlphabet\mathbfss{bold}{OT1}{cmss}{bx}{n}
  \ifAMStwofonts
    \ifCUPmtlplainloaded \else
      \DeclareSymbolFont{UPM}{U}{eur}{m}{n}
      \SetSymbolFont{UPM}{bold}{U}{eur}{b}{n}
      \DeclareSymbolFont{AMSa}{U}{msa}{m}{n}
      \DeclareMathSymbol{\upi}{0}{UPM}{"19}
      \DeclareMathSymbol{\umu}{0}{UPM}{"16}
      \DeclareMathSymbol{\upartial}{0}{UPM}{"40}
      \DeclareMathSymbol{\leqslant}{3}{AMSa}{"36}
      \DeclareMathSymbol{\geqslant}{3}{AMSa}{"3E}

      \let\leq=\leqslant \let\le=\leqslant

    \fi
  \fi
\fi 

\ifCUPmtlplainloaded \else
  \ifAMStwofonts \else 
    \def\upi{\pi}
    \def\umu{\mu}
    \def\upartial{\partial}
  \fi
\fi

\title[The local surface density of disc matter mapped by \Hip]
      {The local surface density of disc matter mapped by \textbfit{Hipparcos}}

\author[J. Holmberg and C. Flynn] {Johan Holmberg$^{1,2}$\thanks{E-mail:
johan@astro.ku.dk} and Chris Flynn$^3$\\
$^{1}$ Astronomical Observatory, NBIfAFG, Juliane Maries Vej 30, DK-2100
Copenhagen, Denmark\\ $^{2}$ Nordic Optical Telescope Scientific Association,
Apartado 474, ES-38 700 Santa Cruz de La Palma, Spain\\
$^{3}$ Tuorla Observatory, V\"ais\"al\"antie
20, FI-21500, Piikki\"o, Finland}

\begin{document}
\maketitle

\begin{abstract}

  We determine the surface density of matter in the disc of the Galaxy at the
solar position using K giant stars. The local space density and luminosity
function of the giants is determined using parallaxes from the \Hip\ satellite;
for more distant giants, observed in a cone at the South Galactic Pole,
distances are determined using intermediate band DDO photometry (which has been
calibrated to the \Hip\ absolute magnitudes). From this sample, we determine
the gravitational potential vertically of the local Galactic disc, by comparing
the number of giant stars observed in the cone with the number expected for
various models of the matter distribution in the disc.  We derive an estimate
of the dynamical disc mass surface density of $56 \pm 6 \Msun\parc^{-2}$, which
may be compared to an estimate of $53 \Msun\parc^{-2}$ in visible matter. For
all gravitating matter (disc + dark halo) we find the total density within 1.1
kpc of the disc midplane to be $74 \pm 6 \Msun\parc^{-2}$. As has been found by
a number of studies, including our own, we find no compelling evidence for
significant amounts of dark matter in the disc.
\end{abstract}

\begin{keywords}
Galaxy: kinematics and dynamics -- Galaxy: structure -- dark matter
\end{keywords}

\section{Introduction}

The measurement of the kinematics and vertical density falloff of a set of
stars in the disc of the Galaxy permit a determination of the dynamical, or
gravitating, local volume density and/or local column density of matter.
Comparison of these quantities with the amount of visible matter has a long
history in astronomy, as any difference between these quantities implies there
remain mass components in the disc which have not yet been directly observed,
i.e. disc dark matter.  The first proposal that there might be a lot of such
unobserved matter dates to Oort (1932,1960).

The European Space Agency's \Hip\ satellite ESA (1997) has had a major
impact on this question, with the consensus now being that there is no
dynamically significant dark matter component in the plane of the disc
(Pham 1997; Cr\'ez\'e et al. 1998; Holmberg \& Flynn 2000). However, these
studies were based on the kinematics and density falloff of nearby A and F
stars. Traditionally, these stars were never regarded as the ideal tracer with
which to work, because they are rather young and it was thought that they would
not be well mixed into the Galactic potential. The studies of A and F stars
observed with \Hip\ now show this view was probably overly cautious; however,
the use of K dwarfs or K giants, being considerably older on average and
certainly well mixed, is a very useful check on the disc mass determinations.

Previous determination of the surface mass density are $48 \pm 9
\Msun\parc^{-2}$, (Kuijken \& Gilmore 1991), $84^{+29}_{-24} \Msun\parc^{-2}$,
(Bahcall, Flynn, \& Gould 1992), $52 \pm 13 \Msun\parc^{-2}$, (Flynn \& Fuchs 
1994), and
$67^{+47}_{-18} \Msun\parc^{-2}$, (Siebert et al. 2003). These numbers give the
total mass of the disc component only, and since the surveys do not extend to
infinite heights above the Galactic plane they include some element of
extra\-polation. To circumvent these two limitations, an alternative way is to
estimate the total gravitating mass (disc + dark halo) within some range in
height above the Galactic plane, preferably limiting the range to the one
actually covered by observations. Kuijken \& Gilmore (1991) estimated
the total mass within 1.1 kpc to $K_{\rm z1.1} = 71 \pm 6 \Msun\parc^{-2}$, and
Siebert et al. (2003) with their shallower survey give $K_{\rm z0.8} =
76^{+25}_{-12} \Msun\parc^{-2}$.

In this paper we return to the K giants in the post-\Hip\ era, and redetermine
the disc surface mass density. The work is a modern revision of the studies by
Bahcall et al. (1992), hereafter BFG and Flynn \& Fuchs
(1994), hereafter FF, which utilize the kinematics and density falloff of K
giants at the South Galactic Pole (SGP) observed by Flynn \& Freeman
(1993). We update these studies by using \Hip\ to greatly improve the
measurements of the local space density and luminosity function of K giants;
and also to improve distance measurements of the K giants at the SGP.  We find
(yet again) that the observational data are very well fit by disc models
containing only the known, visible components: no disc dark matter is required.

In section 2 we describe the selection of the sample from \Hip\ and the
derivation of a new DDO photometry based absolute magnitude scale, as well as
metallicity calibrations for K giants. The cone sample from the SGP is
presented in section 3.  In section 4 we present our model of the disc mass
stratification, and fit the data with and without dark matter in the models,
finding that we can fit the data without invoking disc dark matter. We
summarize and conclude in section 5.

\section{The \textbfit{Hipparcos} Catalogue and selection of sample}

  Our determination of the surface mass density is made from a local sample of
K giants (in a sphere of 100 pc radius around the Sun) and a second sample in a
cone, extending to some 1 kpc vertical to the Galactic plane, at the South
Galactic Pole (SGP). The technique used is elaborated fully in FF, although we
greatly improve on that work by selecting a local sample using the \Hip\
catalogue, improving the determination of the luminosity function and local
density of K giants; and further, by using a \Hip\ calibration of the DDO
photometric system to determine much more reliable distances to the giants in
the cone at the SGP. We also lift the limitations on metallicity used by FF who
only used stars in the interval $-0.5 < $[Fe/H]$ < 0.0$. The basis for the high
metallicity cut at [Fe/H] $= 0.0$ was to exclude young luminous giants from the
sample, on the basis of the best understanding of the age-metallicity relation
at the time; however, the validity of a clear age-metallicity relation has now
been put into question by most modern investigations (Feltzing, Holmberg, \& 
Hurley 2001; Friel et al. 2002; Nordstr\"om et al. 2004), as well as already 
noted
by FF; the evidence for
the high metallicity cut as a means of excluding the youngest stars has
weakened considerably. In any case, with the improved luminosity calibration,
contamination from young, luminous giants is greatly alleviated. The low
metallicity cut at [Fe/H] $= -0.5$ was chosen to exclude kinematically hotter
stars which did not give information on the disk surface density within the
limited vertical range studied; in this new investigation we extend the K giant
study from several hundred parsecs to about 1 kpc above the disc midplane, so
that also the kinematically hotter stars provide information on the Galactic
potential.

\subsection{Local K giants}

\begin{figure} %
\input epsf \center \leavevmode \epsfxsize=0.9 \columnwidth
\epsfbox{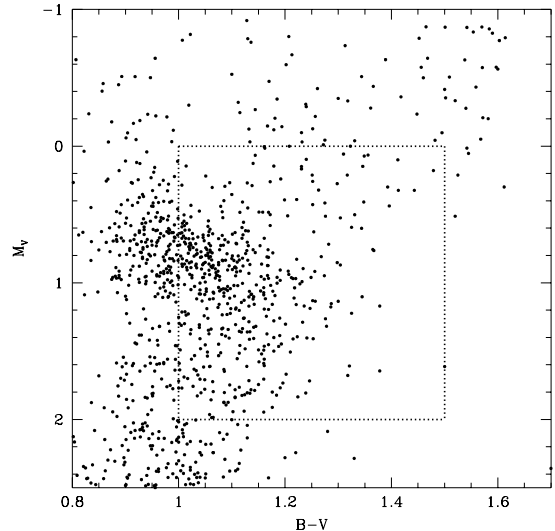}
\caption{CM-diagram of \Hip\ K giants within 100 pc. The dotted box shows the
magnitude and colour limits of the K giant sample used in this paper.}
\label{CMD}
\end{figure}

\begin{figure} %
\input epsf \center \leavevmode \epsfxsize=0.9 \columnwidth
\epsfbox{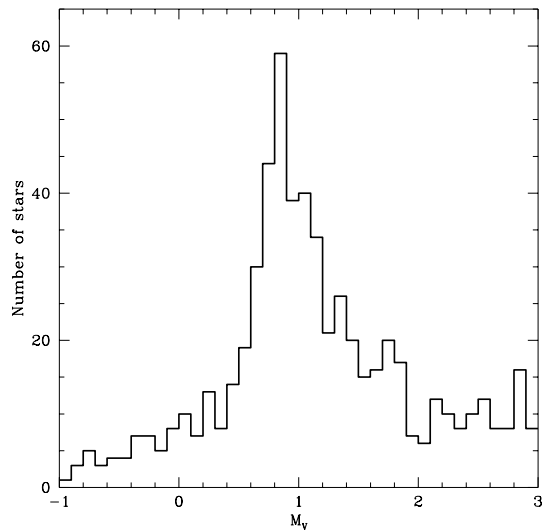}
\caption{Luminosity function of \Hip\ K giants in the local sample within the
colour range $1.0 < B-V < 1.5$.}
\label{lf}
\end{figure}

  The local K giants are drawn from the \Hip\ Survey, that part of the \Hip\
Catalogue intended to be a complete, magnitude limited stellar sample. For
stars of spectral type later than G5 this magnitude limit is $V \leq 7.3 +
1.1~{\rm sin}~|b|$. Figure \ref{CMD} shows the giant region of the \Hip\ colour
magnitude diagram for stars within 100 pc. The giants are bright ($V < 7.5$)
and almost all have intermediate band DDO photometry available.  In
figure~\ref{lf} we show the luminosity function (LF) of the giants in the
colour range $1.0 < B-V < 1.5$. Note that for stars fainter than $M_{\rm V} = 2$, the
sample distance completeness limit moves closer than 100 pc and we make
appropriate corrections to the LF.  The absolute magnitudes are of very high
quality, with errors of order 0.05 mag, i.e. much less than a bin width. The
luminosity function is strongly peaked at $M_{\rm V} \sim 0.85$, the position 
of the
`clump' (i.e. core helium burning) giants on the giant branch.

BFG and FF adopted the absolute magnitude cuts $0.8 < M_{\rm V} < 2.2$ in 
isolating
tracer K giants. The motivation for these cuts were to exclude nearby faint
subgiants, and to avoid luminous giants for which the absolute magnitudes
(which were based on a pre-\Hip\ calibration) were known to be rather poor. The
calibration of the DDO photometric system for the absolute magnitudes of K
giants adopted here does not suffer from deteriorating accuracy with absolute
brightness as the older calibration did, and it is therefore opportune to
change the brighter absolute magnitude cut to $M_{\rm V} = 0.0$. We furthermore
change the lower luminosity limit to $M_{\rm V} = 2.0$. These changes have the 
very
desirable advantage that the absolute magnitude window 
adopted, $0.0 < M_{\rm V} <
2.0$ now brackets the clump stars, rather than cutting into them on the
brighter side of the peak in the LF. Cutting into the LF in the earlier studies
meant that quite careful correction for the Malmquist biases were necessary:
whereas the absolute magnitude window adopted here now renders these
corrections quite straightforward.

We note that although the luminosity cuts now neatly bracket the red giant
clump, the colour cut at $B-V=1.0$ excludes the bluer clump stars. Ignoring the
bluer clump stars potentially introduces bias, as these stars will differ
slightly from their redder counterparts in their age and metallicity
distribution. Unfortunately, we are not at liberty to change this colour limit;
as it is the blue completeness limit of the giants in the cone sample at the
SGP, which was constructed in the pre-Hipparcos era when the colour and
absolute magnitude extent of the clump was poorly known. However, even in the
Hipparcos era, the colour cut turned out to be no real limitation for the
present study. As can be seen in Figure \ref {CMD}, the clump to the blue side
of $B-V = 1.0$ is not as tightly delineated as it is to the red side. The
origin of this effect appears to be the intermixture of young giants, extending
both below and especially above the 'true' clump (consisting of old giants)
 (Girardi 1999). The young giants have a much smaller scaleheight
than the ordinary giants, and would be very scarce in the SGP cone sample. This
is clearly demonstrated by the velocity dispersion of the stars 
with $0.0 < M_{\rm V}
< 2.0$ and $0.8 < B-V \le 1.0$ which is $\sigma_{\rm W}$ = 15.7$\pm0.7$ $\rm
kms^{-1}$, significantly smaller than the redder K giants as will be shown
below. The observational limitations imposed by the pre-\Hip\ studies that the
bluer clump stars must be excluded from the sample seems to have been
expedient.

The sample to 100 pc is volume complete, and allows us to determine the number
density of these giants as $1.10 \pm 0.05 \times 10^{-4}$ per cubic parsec
(recall that they are selected in the windows $0.0 < M_{\rm V} < 2.0$ 
and $1.0 < B-V
< 1.5$). The error in the number density is mainly due to Poisson sampling,
there being 459 giants in the sample; the parallax errors contribute
little. Adopting a typical giant mass of 0.8 \Msun, this implies a mass density
of $0.9 \times 10^{-4}$ \Mden: giants thus represent some 0.2 per cent of the
local disc's stellar content by mass.

\begin{figure} %
\input epsf \center \leavevmode \epsfxsize=0.9 \columnwidth
\epsfbox{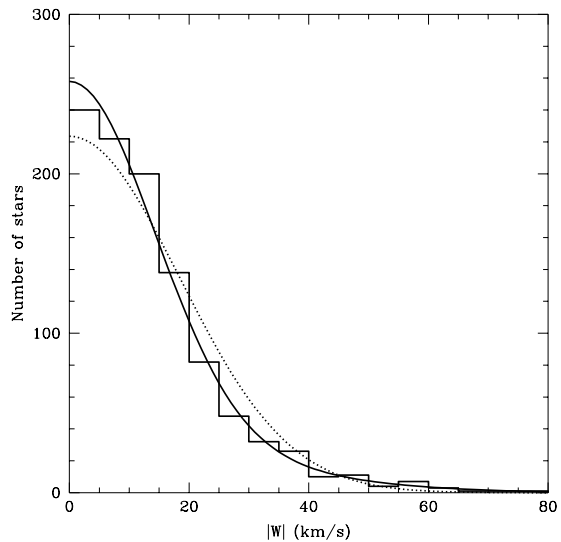}
\caption{Velocity distribution of the local sample of \Hip\ giants (histogram),
and a two component Gaussian fit (solid line). A single component Gaussian is
an insufficiently good fit (dotted line) to the data for the purposes of
computing the density distribution of the stars in a given Galactic potential.}
\label{fw}
\end{figure}

Space velocities for the \Hip\ giants have been calculated as described in
Feltzing \& Holmberg (2000). Radial velocities are available for 438
stars (95 per cent) of our 100 pc volume complete sample. In order to improve
the statistical sampling of the local velocity distribution of the K giants, we
also make use of a larger magnitude limited (rather than distance limited)
sample comprising 1027 stars. The two samples turn out to have very similar
velocity distributions.  Figure~\ref{fw} shows the vertical $W$-velocity
distribution for the large sample of K giants as well as fits with one and two
component Gaussians. The fitted velocity distributions have been corrected for
the small contribution from measurement errors in distance, radial velocity and
proper motion determination. With a single isothermal component, the velocity
dispersion $\sigma_{\rm W}$ is 18.3$\pm0.4$ $\rm kms^{-1}$, but a much improved 
fit
is obtained with two components with velocity dispersions of 14.0$\pm0.8$ and
28.3$\pm2.9$ $\rm kms^{-1}$ and a relative density from the hot component of
0.23$\pm0.08$. For the small sample the corresponding numbers are:
18.2$\pm0.6$, 14.0$\pm1.1$, 30.0$\pm4.8$ $\rm kms^{-1}$ and a relative density
of 0.19$\pm0.10$.

The dispersions and relative densities were found using a maximum likelihood
decomposition routine, which also give the confidence intervals of the fitted
parameters. For the determination of the surface density of matter, the
individual components of the velocity dispersion fit matter little; it is their
combined distribution which is used as a smooth fit to the observed velocity
distribution. The other components of the space velocity distribution are :
$\sigma_{\rm U} = 34 \rm kms^{-1}$, $\sigma_{\rm V} = 23 \rm kms^{-1}$, and 
$V_{\mathrm
lag} = -18 \rm kms^{-1}$.

\begin{figure} %
\input epsf \center \leavevmode \epsfxsize=0.9 \columnwidth \epsfbox{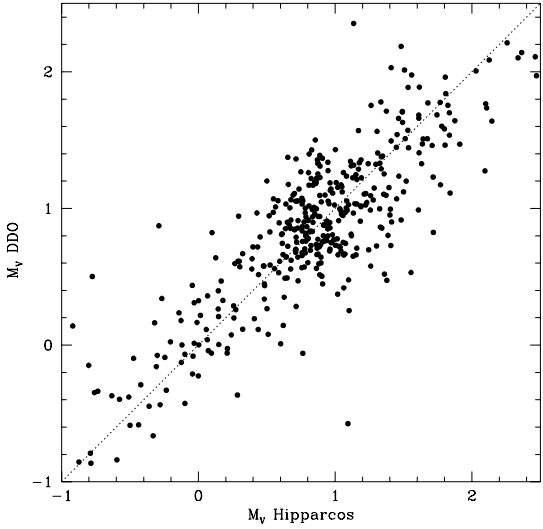}
\caption{Comparison between the fitted DDO absolute magnitudes and \Hip\
magnitudes, as determined from parallaxes.}
\label{mvcal}
\end{figure}

\subsection{DDO absolute magnitude calibration}

Starting from the high quality \Hip\ determinations of absolute magnitude for
nearby K giants, we have derived a new calibration of giant luminosity in the
DDO intermediate band photometric system. We use all three DDO colours (C4142,
C4245 and C4548) to fit for metallicity as well as absolute magnitude in the
new calibration. The main difference between this determination and a similar
calibration by H{\o}g \& Flynn (1998), is that they restricted their study
to metal rich giants ([Fe/H]$ > -0.5$), whereas our new calibration is valid
for the whole metallicity range of our study. DDO measurements were found in
Mermilliod \& Nitschelm (1989) for 419 stars in our local sample for
the luminosity range $-2 < M_{\rm V} < 3$. For this dataset we fitted a 
third-order
polynomial containing all combinations of the three DDO colours using the
downhill simplex amoeba routine (Press et al. 1992). The resulting calibration
equation is: \\


\noindent $\rm M_V = 55.14-32.18c_{1}+15.90c_{2}-48.91c_{3}-82.34c_{1}^{2} \\
+5.05c_{2}^{2}+88.82c_{3}^{2}+20.54c_{1}c_{2}+87.62c_{1}c_{3}+12.85c_{2}c_{3}\\
+49.69c_{1}^{3}+23.36c_{2}^{3}+135.99c_{3}^{3}+9.52c_{1}^{2}c_{2} -
79.24c_{1}^{2}c_{3} \\
-53.58c_{2}^{2}c_{1}-55.47c_{2}^{2}c_{3}-189.84c_{3}^{2}c_{1} +
97.40c_{3}^{2}c_{2}+84.97c_{1}c_{2}c_{3}$ \\

\noindent
where $c_{1} = C4548, c_{2} = C4245$, and $c_{3} = C4142$. Figure~\ref{mvcal}
shows the relation between the fitted DDO absolute magnitudes and the
calibrators from \Hip.\ The scatter of the fit is 0.35 mag and note that the
calibration is valid in the ranges: $1.10 < C4548 < 1.40, 0.80 < C4245 < 1.30,
-0.10 < C4142 < 0.40, {\rm and} -2 <M_{\rm V} < 3$.

To asses the precision of the fitting process, a number of bootstrapped
DDO-\Hip\ catalogues were created. These catalogues were then used as input to 
the
fitting routine to produce alternative calibration equations. Finally the
original catalogue was fed into this ensemble to produce DDO magnitudes which
could be compared to the original ones. The resulting mean dispersion of only
0.07 magnitudes shows the high stability of the derived absolute magnitudes to
variations of the set of calibration stars.

\begin{figure} %
\input epsf \center \leavevmode \epsfxsize=0.9 \columnwidth
\epsfbox{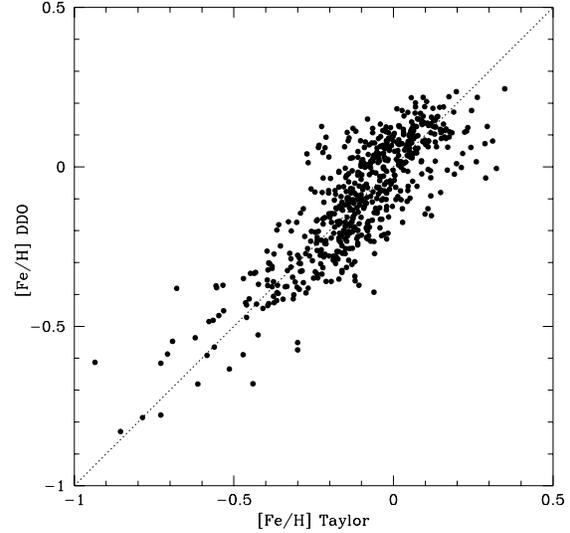}
\caption{Comparison between the fitted DDO-photometry metallicities and
calibrating spectroscopic metallicities.}
\label{fehcal}
\end{figure}

\subsection{DDO metallicity calibration}

In a similar way to the new DDO $M_{\rm V}$ calibration, we have also 
determined a
new DDO [Fe/H] calibration. Here we used in addition to the DDO photometry from
Mermilliod \& Nitschelm (1989) the catalogue of giants with
spectroscopic metallicities of Taylor (1999). For 597 stars found in
both sources we fitted a relation of the same functional form as the $M_{\rm V}$
calibration. The resulting calibration equation is: \\

\noindent 
$\rm [Fe/H] = 6.30-5.67c_{1}+6.05c_{2}-13.27c_{3}-3.65c_{1}^{2} \\
-2.42c_{2}^{2}+29.41c_{3}^{2}-4.36c_{1}c_{2}+9.25c_{1}c_{3}+1.51c_{2}c_{3} \\
+2.93c_{1}^{3}+3.02c_{2}^{3}-83.39c_{3}^{3}-0.46c_{1}^{2}c_{2} +
7.66c_{1}^{2}c_{3} \\
-0.12c_{2}^{2}c_{1}-13.53c_{2}^{2}c_{3}-26.76c_{3}^{2}c_{1} +
58.76c_{3}^{2}c_{2}-4.41c_{1}c_{2}c_{3}$ \\

\noindent
where $c_{1} = C4548, c_{2} = C4245$, and $c_{3} = C4142$, as
before. Figure~\ref{fehcal} shows the relation between the fitted DDO
metallicities and those from the spectroscopic sources. The scatter of the fit
is 0.11 dex and the calibration valid in the ranges: $1.14 < C4548 < 1.43, 0.77
< C4245 < 1.33, 0.08 < C4142 < 0.45$, and $-0.93 < $ [Fe/H] $ < 0.34$.

\section{SGP K giants}

BFG and FF analyzed the sample of K giants at the SGP of Flynn \& Freeman
(1993) for which absolute magnitudes and hence distances were determined
using DDO photometry, using a calibration which much predated the \Hip\
results. In this paper we reanalyze the sample, now with revised K giant
absolute magnitudes via the improved, \Hip\hspace{-1mm} based calibration of
giant absolute magnitudes described in the previous section.

\begin{figure}   
\input epsf \center \leavevmode \epsfxsize=0.9 \columnwidth \epsfbox{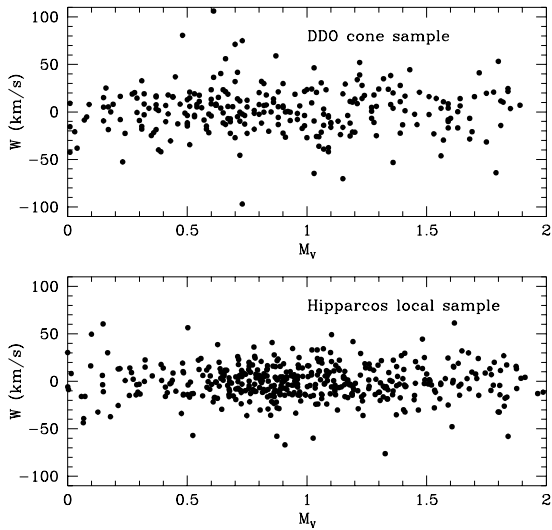}
\caption{The vertical $W$ velocities of the giants as a function of absolute
magnitude in the local and cone samples.}
\label{mvw}
\end{figure}

\begin{figure}   
\input epsf \center \leavevmode \epsfxsize=0.9 \columnwidth \epsfbox{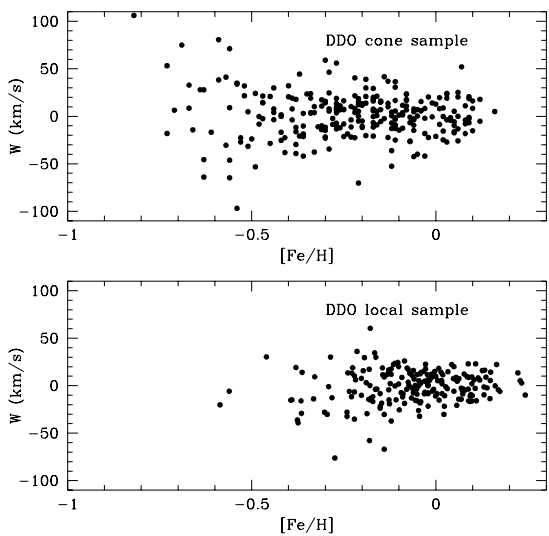}
\caption{The vertical $W$ velocities as a function of metallicity in the local
and cone samples.}
\label{fehw}
\end{figure}


The first sample we analyze (the `HD sample') consists of all K giants at the
SGP in the Flynn \& Freeman (1993) catalogue to a limiting visual
magnitude of $V = 9.2$ and for which $0.0 < M_{\rm V} < 2.0$ 
and $1.0 < B-V < 1.5$,
resulting in 139 K giants in a 430 square degree region.


In addition to the HD stars in the 430 square degree region at the South
Galactic Pole, Flynn \& Freeman (1993) observed candidate K giants in a
40 square degree region from the catalogue of Eriksson (1978). This
catalogue extends to $V = 11.0$. Candidate giants were selected in the colour
range $0.95 < B-V < 1.55$ and DDO photometry and radial velocities obtained.

Flynn \& Freeman also observed stars with $B-V > 0.7$ from their own plate
material obtained with a 6 inch Zeiss camera mounted on the 9 inch Oddie
refractor at Mount Stromlo, reaching to $V = 11.2$. The Oddie survey covered
100 square degrees.

The giants from these two sources will be referred to as the Eriksson and Oddie
samples. Neither of these two samples were included in the disc surface mass
density analyses of BFG or FF; this was primarily because the poorer absolute
magnitudes of these fainter giants, as determined by DDO photometry, introduced
error into the determinations which outweighed the advantage of better Poisson
statistics. This expedient, while appropriate at the time, can now be dropped
because of two major improvements. Firstly, we now have a greatly improved
absolute magnitude calibration of the DDO photometric system for the giants
which is tied to the very high precision \Hip\ parallaxes. Secondly, we can
utilize the Tycho-2 catalogue (H\o g et al. 2000) to directly determine the
completeness of these samples (and the HD sample for that matter).  The deeper
samples improve the disc mass determination by extending the height to which
the giants probe the Galactic potential to about 1000 pc above the disc,
whereas the HD sample does not probe beyond 500 pc. The Eriksson and Oddie
catalogues add 212 K giants to the sample, again selected in the range $0.0 <
M_{\rm V} < 2.0$ and $1.0 < B-V < 1.5$.


Due to the availability of the Tycho-2 catalogue, it is now possible to
complement the HD, Eriksson and Oddie samples with proper motions to obtain
full space velocities. The Tycho-2 catalogue consists of $\sim$2.5 million
stars, and is 99 per cent complete to $V = 11$. This is well borne out by our
sample: where 350 stars (99.7 per cent) have measured proper motions in the
Tycho-2 catalogue. Together with the radial velocities from Flynn \& Freeman
(1993) we obtain full space velocities for 264 stars in our cone
sample. In general the velocity 
dispersions ($\sigma_{\rm U}, \sigma_{\rm V}, \sigma_{\rm W},
V_{\mathrm lag}$) = (41, 32, 25, $-23$) $\rm kms^{-1}$ of the cone sample are
larger than the corresponding ones in the local sample in accordance with the
increasing contribution of the thick disc above the Galactic
plane. Figure~\ref{mvw} shows the $W$ velocities as a function of absolute
magnitude for the DDO cone sample compared to the \Hip\ local
sample. Figure~\ref{fehw} shows the corresponding distribution of $W$
velocities as a function of metallicity for the DDO cone sample compared to the
DDO local sample.

We also use the Tycho-2 magnitudes and colours to check the completeness of the
SGP cone sample. When using this information from Tycho-2 it is very important
that the full transformations (rather than the first order approximations) from
the Tycho instrumental $V_{\rm T}$ and $B_{\rm T}$ colours into the standard 
Johnson
$V$ and $B-V$ are used. This approach which was used in obtaining the $B-V$
colours given in the Hipparcos catalogue is as described in section
1.3 and Appendix 4 of Volume 1 of the catalogue. We note that use of the first
order relations can lead to biases of up to 0.05 in the $B-V$ colour, which is
insufficiently accurate for testing sample completeness.  For the HD catalogue,
which is reported by Flynn \& Freeman (1993) as complete to $V = 9.2$ and
extending to about $V = 10.5$ with decreasing completeness, we confirm that the
sample is indeed complete to $V = 9.2$. For the Eriksson and Oddie catalogues,
we find that they are indeed complete to $V = 11.0$.

\section{Surface mass density determination}

\begin{figure} %
\input epsf \center \leavevmode \epsfxsize=0.9 \columnwidth
\epsfbox{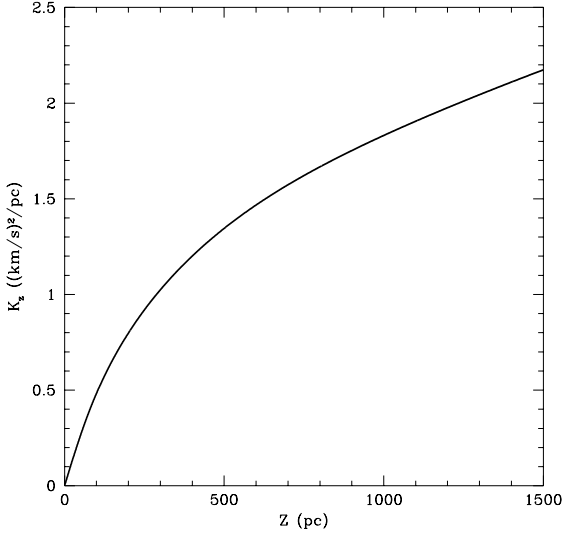}
\caption{The $K_{z}$ force law resulting from the basic mass model.}
\label{kz}
\end{figure}

We adopt a disc model similar to the one used in Holmberg \& Flynn
(2000) to analyze the data.  The basic model is shown in Table 1 (and
described in section 3) of Holmberg \& Flynn (2000): it is of the
type introduced by Bahcall (1984a,b,c), in which the disc is represented by a
set of massive, kinematically isothermal components, tracing young stars, old
stars, stellar remnants and gas.  The Poisson-Boltzmann equations are solved
simultaneously and the density falloff of each component computed. The
difference between the model used here and the old one is that a thick disc
component is explicitly included with a local density of $0.007
\Msun\parc^{-3}$, and a velocity dispersion of 37 $\rm kms^{-1}$. The density
and velocity dispersions of the appropriate thin disc components are reduced
accordingly, to keep the total local density and velocity distributions in line
with the bounds determined by Holmberg \& Flynn (2000). Some
properties of or basic mass model are as follows: local mass density $\rho_{0}
= 0.102 \Msun\parc^{-3}$, local {\it disc} mass surface density $\Sigma = 52.8
\Msun\parc^{-2}$, and total local mass surface density within 1.1 kpc of the
Galactic mid-plane $\Sigma_{\rm z1.1} = 70.6 \Msun\parc^{-2}$. The resulting
$K_{\rm z}$ force law of the model is shown in Figure~\ref{kz}.

\begin{figure} %
\input epsf \center \leavevmode \epsfxsize=0.9 \columnwidth
\epsfbox{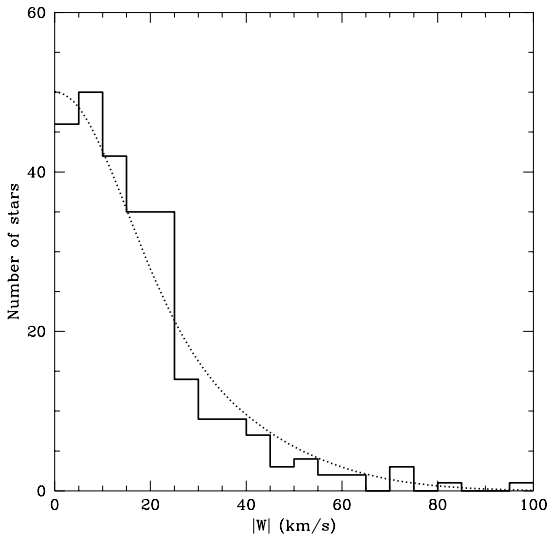}
\caption{Velocity distribution of the SGP cone sample of giants (histogram),
compared to the expected velocity distribution (dotted line). This is calculated
from the mass model and the local velocity distribution for the same height
distribution as the observed sample.}
\label{fwcone}
\end{figure}

Figure~\ref{fwcone} shows a comparison between the measured velocity
distribution in the cone sample and the expected distribution, for which
$\sigma_{\rm W} = 26 \rm kms^{-1}$. The expected distribution is computed from 
the
combination of the mass model and the velocity distribution in the plane of the
disc (as outlined in detail in FF). The good agreement
between the measured and calculated distributions is an indication that the
local and the cone sample are members of the same tracer population and well
suited for the determination of the mass density.

\begin{figure} %
\input epsf \center \leavevmode \epsfxsize=0.9 \columnwidth
\epsfbox{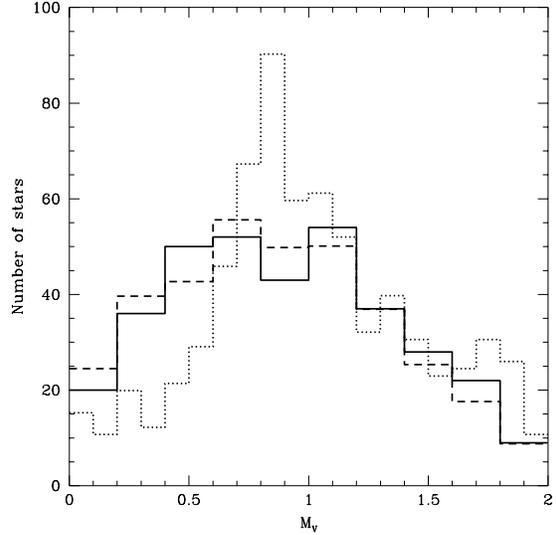}
\caption{Comparison between the local \Hip\ luminosity function (LF)
(normalized to the number of stars in the cone sample, dotted line) and the one
from the DDO cone sample (full line). When the local \Hip\ LF is convolved by
the DDO cone sample measurement uncertainties and selection effects, the result
is the dashed histogram. The local LF is thus found to be a good match to the
LF of the cone giants.}
\label{lfcomp}
\end{figure}

Figure~\ref{lfcomp} shows a comparison between the local and the cone
luminosity functions (LFs). The luminosity function of the cone K giants
appears to be quite consistent with being drawn from the same LF as the local K
giants. This is in consideration of the much larger typical error in the
absolute magnitudes (0.35 mag in the cone compared to 0.05 mag for the local
giants), and the fact that the cone sample is magnitude rather than volume
limited. Under the assumption that the LFs are the same, then from measured
local density of the K giants and the predicted density falloff of the giants
in a particular model of the disc mass, we can compute the number of giants
which should appear in the cone survey. This can be compared to the actual
number of observed giants and the model evaluated. In the manner of FF we ran a
series of Monte Carlo simulations of observations of stars in the cone samples,
taking into account observational uncertainties and selection effects, in order
to compute the expected number of giants in the cone for various mass models of
the disc. The error sources include the uncertainty in the local density of the
tracer stars and the uncertainty in the fitted velocity distribution of the
local sample. This was determined by making a series of bootstrapped velocity
catalogues, and then fitting a two-component Gaussian in the same way as for
the observed velocity sample. Finally the goodness of the fit to the
observations were determined using a standard $\chi^{2}$ statistic.

\begin{figure} %
\input epsf \center \leavevmode \epsfxsize=0.9 \columnwidth
\epsfbox{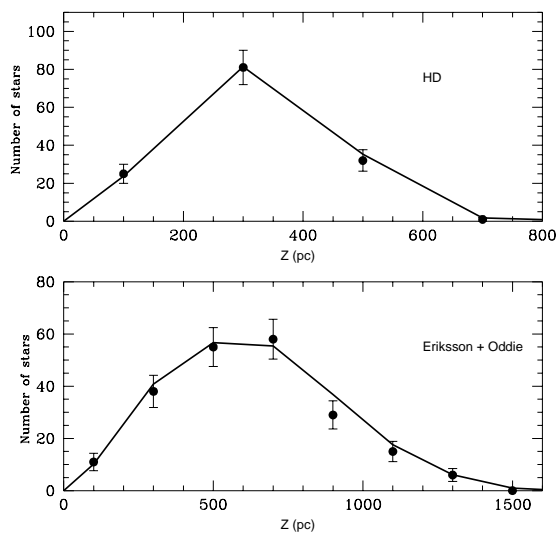}
\caption{Predicted (solid curve) versus observed numbers of giants in the cone
at the SGP for the HD and Eriksson+Oddie samples. The predictions are for the
disc model containing visible matter only, shown as a function of $Z$ height
above the Galactic mid-plane. The no-disc-dark-matter model is an excellent fit
to the data in all parts of the survey.}
\label{zcounts}
\end{figure}

We find that the basic disc mass-model gives a very good description of the
vertical distribution of stars in the cone samples. For the HD-sample it
predicts 142 stars compared to the 139 observed and for the Eriksson+Oddie
sample it predicts 224 stars compared to the observed 212. Figure~\ref{zcounts}
shows the observed vertical distribution of stars above the Galactic plane
compared to the predicted one. The fit is excellent. Clearly, the visible
components of the disc are entirely sufficient to explain the data: disc dark
matter is obviated yet again. Nevertheless, to exactly reproduce the
distribution of stars in the cone sample a small amount of additional disc
matter can be added. This is achieved with a total disc surface density of
$56 \pm 6 \Msun\parc^{-2}$. The total amount of gravitating mass to 800pc and
1.1 kpc is determined to be $65 \pm 6 \Msun\parc^{-2}$ and $74 \pm 6
\Msun\parc^{-2}$ respectively.

Recently, Korchagin et al. (2003), have made a determination of the surface mass
density of the local disc using K giants in the Hipparcos catalogue. Their sample
utilizes mainly first ascent giants, rather than the clump giants which we have
used here. They have selection criteria by colour and absolute magnitude which
are intended to isolate the older giants. Their sample extends to about 350 pc
from the Galactic plane, because they only use giants with Hipparcos
parallaxes.  They derive a surface mass density $K_{\rm z0.35} = 42 \pm 6
\Msun\parc^{-2}$ within 350 pc of the mid-plane. This is in excellent agreement
with our model, which gives  $K_{\rm z0.35} = 41 \Msun\parc^{-2}$.

\section{Conclusions}

We have reanalyzed the sample of K giants studied by Flynn \& Freeman
(1993), Bahcall, Flynn \& Gould (1992) and Flynn \& Fuchs (1994) 
for determining the mass density and surface mass density of the local Galactic
disc. The reanalysis incorporates the absolute magnitudes for nearby K giants
measured by the \Hip\ satellite, and a calibration of absolute magnitudes for
distant giants.  Our results confirm that there is no significant missing
component in the mass inventory of the local Galactic disc. We derive a surface
mass density of material at the Solar position of $K_{\rm z1.1} = 74 \pm 6
\Msun\parc^{-2}$ within 1.1 kpc of the mid-plane, with $56 \pm6 
\Msun\parc^{-2}$ as the total disc contribution, compared to 53
$\Msun\parc^{-2}$ in visible material.  This is in excellent agreement with the
analysis of K dwarfs, for which Kuijken \& Gilmore (1991) estimated the
total mass within 1.1 kpc as $K_{\rm z1.1} = 71 \pm 6 \Msun\parc^{-2}$.

\section*{Acknowledgments}

CF thanks the Academy of Finland and the ANTARES program for its support of
space based research; and also support by the Beckwith Trust. JH thanks the
Carlsberg Foundation and the Council of the Nordic Optical Telescope Scientific
Association. Lennart Lindegren is thanked for valuable assistance. The paper is
based on data obtained with the European Space Agency's \Hip\ astrometry
satellite. We have made extensive use of the Simbad and VizieR stellar data
bases, at the Centre de Donn\'ees astronomiques de Strasbourg, and of NASA's
Astrophysics Data System. The plate material for one of the surveys was
obtained at Mount Stromlo with the Oddie telescope, which dates from 1889 and
which was destroyed, along with all the other telescopes on the mountain, in
the bushfires of January 2003; happily there are plans afoot to restore this
wonderful instrument. CF acknowledges the nightly assistance of itinerant
Stromlo wildlife in keeping him alert while guiding plate exposures at the
Oddie.
 
{} 

\end{document}